\documentstyle[manuscript,aps]{revtex}

\begin{document}
\title{Bessel-Gauss beam optical resonator}
\author{J. Rogel-Salazar, G. H. C. New, S. Ch\'{a}vez-Cerda$^{\star }$}
\address{Laser Theory Group, The Blackett Laboratory, 
Imperial College, London SW7 2BW}
\maketitle

\begin{abstract}
In a simple picture, a Bessel beam is viewed as a transverse standing wave
formed in the interference region between incoming and outgoing conical
waves. Based on this interpretation we propose an optical resonator that
supports modes that are approximations to Bessel-Gauss beams. The Fox-Li
algorithm in two transverse dimensions is applied to confirm the conlcusion.
\end{abstract}

\vspace*{3.5in}

\noindent {\large PACS:}

{\bf Lasers}: application, 42.62; general theory of, 42.55.A; optical
systems, 42.60;

types of, 42.55.

{\bf Resonators}: lasers, 42.60.D; optical, 42.79.Y

\newpage

In view of the many potential applications of Bessel beams (BBs), there is
strong motivation to design a laser with a Bessel beam output. For example,
a high-power beam with minimal diffractive spreading could be used for
precise machining and cutting, while BBs have been identified as possible
conduits in atom optics and micro-manipulation experiments \cite
{tremblay,arlt,sabino}. With this kind of applications in mind the focusing
characteristics of BBs have been studied in detail, and it has been shown
that very sharp and intense beams can be produced \cite{chavezoc}.

Laser resonators with BB modes have been studied previously, but the designs
have lacked inherent simplicity and the concepts have consequently not been
fully exploited \cite{durninpat,uehara,onae,jab,belyi,perti}. With the
exception of the model in Ref. \cite{perti} where the diffractive features
of Bessel-Gauss beams were applied, all previous designs for laser
resonators with BB modes were based on Durnin's scheme in which a BB was
created with an annular slit followed by a lens \cite
{durninpat,uehara,onae,jab,belyi}. Since the overlap between the mode and
the gain medium is limited in these designs, their efficiency is generally
low.

The structure of BBs can be interpreted in several different ways. Durnin
and Eberly pointed out that a BB can be regarded as the superposition of an
infinite set of plane waves whose wave vectors lie on the surface of a cone 
\cite{durnin}. Alternatively, a BB can be decomposed into a pair of conical
waves, one slanting inwards towards the axis, and the other expanding away
from it \cite{chavez1}. Mathematically, the two components are represented
by Hankel functions which are fundamental solutions of the Helmholtz
equation.

In this paper, we study the properties of a simple resonator design that
supports modes with BB characteristics. The mode profiles turn out in fact
to be close to Bessel-Gauss beams and we offer an explanation for this
result.

Any practical BB is invariably of {\it finite} extent \cite{chavez1}, and
the region in which the constituent conical waves overlap, and the beam
retains its essential integrity, is shown dark shaded in the simple
geometrical construction of Fig. \ref{fig:conicregion}. We will refer to
this conical region as the cone of existence of a BB. The corresponding
profile evolution obtained by numerical solution of the Helmoltz equation is
shown in Fig. \ref{fig:bb_phase}, where we can observe the propagation of
the finite BB and the behaviour of the associated wavefronts. Notice that
the cone of existence of the BB is clearly apparent, and that outside this
cone, the wavefronts are practically conical.

The wavefronts of Gaussian beam modes are spherical so a cavity formed by
spherical mirrors supports a Gaussian beam \cite{siegman}. By analogy, since
the constituent wavefronts of a BB, {\it i.e.} the Hankel waves, are
conical, it is logical to seek to support a BB mode using conical mirrors.
One possibility would be to employ two identical conical mirrors, but in
this case the BB would be formed only inside the resonator \cite{jab}. The
alternative is to form a cavity with a conical mirror and a plane mirror
(which forms a virtual image of the former) in such a case both the cavity
mode and the output are BBs.

Two closely related BB resonator designs based on this principle are shown
in Figs. \ref{fig:resonators}a and \ref{fig:resonators}b. While the first
uses a conical mirror, this is replaced in Fig. \ref{fig:resonators}b with a
refractive axicon backed by a plane mirror. Mechanically it is simpler to
manufacture {\it convex} conical surfaces, so this second option may be more
viable in practice. Nevertheless, a Fresnel conical mirror \cite{fuji} can
be used as an option for the concave configuration. A further alternative
for solid-state media might be to shape the end of the rod into a cone and
deposit a reflecting layer on this surface.

We note that conical mirrors with a $90^{\circ }$ vertex angle have
sometimes been used in the past in preference to plane mirrors to facilitate
cavity alignment in high-power unstable resonators \cite{geoff}. Also, $%
90^{\circ }$ conical mirrors have been used to enhance efficiency of some
laser systems by improving the pumping uniformity \cite{kuhnle}.

A geometrical analysis shows that to maximise the gain in the cavity volume,
the length of the cavity must be equal to the maximum propagation distance
of a finite Bessel beam $Z_{max }$ (see Fig. \ref{fig:conicregion}). In
terms of the resonator parameters $L=Z_{max }=A/(2\tan \theta )$ where $A$
is the radius of the conical mirror or the axicon and the angle $\theta $ is
related to the transverse $k_{r}$ and longitudinal $k_{z}$ components of BB
the wave vector through $\tan \theta =k_{r}/k_{z}$. The conical mirror angle 
$\alpha =\theta /2$, while for the refractive axicon $\beta =\theta /(n-1)$,
where $n$ is the refractive index (see Figs. \ref{fig:resonators}).

Referring to Fig. \ref{fig:resonators}a, we notice that, within the cavity,
each plane wave component in the conical wave will ``see'' a finite
Fabry-Perot resonator with mirrors of sides $A$ and separated by a distance $%
\sim 2L$. It is known that the corresponding lowest-order mode of a
plane-plane cavity has a bell-shaped amplitude profile \cite{foxli,siegman};
this suggests that the modes will be the superposition of the Hankel conical
waves modulated by a bell-shaped function. The modes of the cavity will
therefore have similar features to Bessel-Gauss beams \cite{gori}.

We have used the Fox-Li method to obtain the fundamental field mode for the
resonator \cite{foxli,siegman}. To account for losses at the open sides of
the resonator we propagated the field within the cavity by solving the
Helmholtz equation with absorbing boundary conditions \cite{chavez2}.

The geometrical values used for the simulations were $A=40,L=35$ in
normalised units and the angle of the axicon mirror was computed by the
relation given above. Our data are related to a real physical system whose
dimensions are 40 cm length, 2 cm diameter. The axicon refractive index $%
n=2.4028$ is that of ZnSe at 10.6 $\mu $m. With these values, the angle of
the required axicon is computed to be approximately $0.5^{\circ }$.

Instead of the Fox-Li eigenvalue parameter $\gamma $, we have chosen to
analyse the behaviour of the relative power loss per round trip, ${\cal L}
_{k}=(I^{k}-I^{k+1})/I^{k}$, where $I^{k}=\int |u(x,y)|^{2}dxdy$ is the
integrated intensity of the field at the $k$-th transit. There is a simple
relation between the relative losses and the eigenvalue of the modes, namely 
${\cal L}=1-|\gamma |^{2}$. With this relation the threshold gain is $
G=1/|\gamma |^{2}=1/(1-{\cal L})$.

We have performed simulations for a wide range of values and a variety of
initial conditions, including uniform plane waves, Gaussian and
super-Gaussian beams with different widths and irregular profiles with
random positions in the transverse input plane. A normal sampling grid was $%
512\times 512$ but double density sampling was used to check convergence of
the results. As expected, the outcome was a BB modulated by a bell-shaped
function. Irrespective of the initial condition, the field always converged
to the zero-order mode of the cavity with the expected profile and radial
frequency features imposed by the geometrical parameters. We also computed
the mode of the flat-flat cavity discussed above to obtain the bell-shaped
modulation and fitted a Gaussian profile. When the computed modes for the
resonator with the conical mirror were compared with a Bessel-Gauss beam,
the differences were negligible.

Higher order Hankel waves have an azimuthal phase factor, $\exp (im\varphi )$
with $m$ integer and zero field on axis. To induce this kind of field, we
used initial conditions of the form $r\exp (-r^{2}/w_{0}^{2})\exp
(im_{0}\varphi )$, $w_{0}=5$. Although we were able to produce higher order
BB modes, we only obtained modes with $m=1$ and $m=2$, regardless of the
initial topological charge $m_{0}$ and the value of $w_{0}$. The reason is
that the transverse dimensions of the cavity were fixed and so the
higher-order modes have higher losses and could not be sustained by the
cavity. Typical profiles of zero and first order modes are shown in figures 
\ref{fig:mode}a and \ref{fig:mode}b.

The behaviour of the losses as a function of round trip is shown in figure 
\ref{fig:loss}. We observe that there is a transient state around which the
mode oscillates with high losses. This state is however unstable and the
system accomodates to a stable mode with lower losses. During the transient,
the field profile agrees with the induced $m_{0}$ mode that later decays to
the transverse Bessel mode profile with $m=0,1$ or $2$. Once the mode is
stabilised, the losses remain constant and, for the configuration studied,
remained under 4\% for the zero order mode. However, the losses were higher
for the modes with $m=1,2$, although they still were under 7\% and 12\%,
respectively. Despite the relative high losses, the modes are very well
defined.

Spherical mirrors can be used to minimise the losses due to diffraction and
to create true Bessel-Gauss beams. To this end, curvature was included in
the conical mirror in Ref. \cite{perti}. However, the same effect can be
achieved much more simply in the configuration proposed here, by replacing
the flat mirror with a spherical mirror. 

To test the stability of the higher modes, the initial condition was
modified by adding white noise with magnitude 1\% of the maximum initial
amplitude. Azimuthal features could not be sustained by the cavity under
these conditions and the final result was a zero order mode. The physical
reason is that the spectrum of the noisy profile contains higher frequencies
increasing the spillover at the open sides of the cavity.


In conclusion, we have demonstrated the possibility of creating a resonator
whose modes and output are close to Bessel-Gauss beams. The Gaussian-like
modulation is inherited from the finite Fabry-Perot resonator mode. Our
results show that this kind of configuration can also support higher order
modes when excited with an initial condition carrying azimuthal phase and
zero-valued on axis. However, these modes are not stable, and small
irregular variations in the amplitude profile, produced for instance by
spontaneous emissions, can induce their decay to the zero order mode. The
design presented here is simple to construct and makes good usage of the
volume of the cavity, enhancing power performance. Also, it can be used for
solid-state lasers for which the conic mirror can be machined at one of the
rod ends. The design can be extended to unstable resonators by changing the
flat mirror into a convex spherical mirror.

The authors acknowledge fruitful discussion with J. C. Guti\'{e}rrez-Vega
and P.Muys, who also gave the experimental support.

This work has been partially supported by CONACyT and EPSRC grant number
GR/N37117/01.

$^{\star }$On leave from Instituto Nacional de Astrof\'{\i}sica, Optica y
Electr\'{o}nica, Apdo. Postal 51/216, Puebla, Pue. M\'{e}xico 72000.

\newpage
\begin{center}
{\Large Figure Captions}
\end{center}

\begin{enumerate}
\item  \label{fig:conicregion} Picture showing the finite region where
conical waves overlap and produce a Bessel beam (dark shading). Outside this
region the conical waves continue propagating (light shading).

\item  \label{fig:bb_phase}Intensity and phase propagation of a finite
Bessel beam.

\item  \label{fig:resonators}Cavity configurations for Bessel beams, a) with
reflective axicon, b) equivalent configuration with refractive axicon.

\item  \label{fig:mode}Cavity modes obtained numerically of the proposed
resonator. a) Zero order mode; b)First order mode.

\item  \label{fig:loss}Representative curves of the relative loss per
transit.
\end{enumerate}


\begin{references}
\bibitem{tremblay}  M. Florjanczyk, R. Tremblay, Optics. Comm. {\bf \ 73},
448 (1989).

\bibitem{arlt}  J. Arlt, M. J. Padgett, Opt. Lett. {\bf 25}, 191 (2000); 
  
\bibitem{sabino} S.  Ch\'{a}vez-Cerda, E. Tepich\'{\i}n, M. A.
  Meneses-Nava, G. Ram\'{\i}rez, J.  M. Hickmann, Opt. Express {\bf 3}
  524-529 (1998).

\bibitem{chavezoc}  S. Ch\'{a}vez-Cerda, G. H. C. New, Optics Comm. {\bf 181}%
, 369-377 (2000).

\bibitem{durninpat}  J. Durnin, J. H. Eberly, US Patent 4 887 885, 1989.

\bibitem{uehara}  K. Uehara, H. Kikuchi, Appl. Phys. {\bf B 48}, 125-129
(1989).

\bibitem{onae}  A. Onae, T. Kurosawa, Y. Miki, E. Sakuma, J. Appl. Phys. 
{\bf 72}, 4529-4532 (1992).

\bibitem{jab}  J. K. Jabczyoski, Optics Comm. {\bf 77}, 292-294 (1990).

\bibitem{belyi}  V. N. Belyi, N. S. Kazak, N. A. Khilo, Optics Comm. {\bf 162%
}, 169-176 (1999).

\bibitem{perti}  P. P\"{a}\"{a}kk\"{o}nen, J. Turunen, Optics Comm. {\bf 156}
, 359-366 (1998).

\bibitem{durnin}  J. E. Durnin, J. J. Miceli and J. H. Eberly, Phys. Rev.
Lett. {\bf 58}, 1499-1501 (1987).

\bibitem{chavez1}  S. Ch\'{a}vez-Cerda, J. Mod. Opt. {\bf 46}, 923-930
(1999).

\bibitem{foxli}  A. G. Fox, T. Li, Bell Sys. Tech. J. {\bf 40}, 453-488
(1961).

\bibitem{siegman}  A. E. Siegman, {\it Lasers}, University Science Books,
Sausalito CA,1986.

\bibitem{fuji}  S. Fujiwara, J. Opt. Soc. Am. {\bf 51}, 1305 (1961).

\bibitem{geoff}  C. A. Moreira, G. H. C. New, Appl. Phys. B {\bf 56},
373-378 (1993).

\bibitem{kuhnle}  G. Kuhnle, G. Marowsky, G. A. Reider, Appl. Opt. {\bf 27},
2666-2670 (1988).

\bibitem{gori}  F. Gori, G. Guattari, C. Padovani, Optics Comm. {\bf \ 64},
491-495 (1987).

\bibitem{chavez2}  S. Ch\'{a}vez-Cerda, M. A. Meneses-Nava, J. M. Hickmann,
Optics Lett. {\bf 25}, 83 (2000).

\end{references}
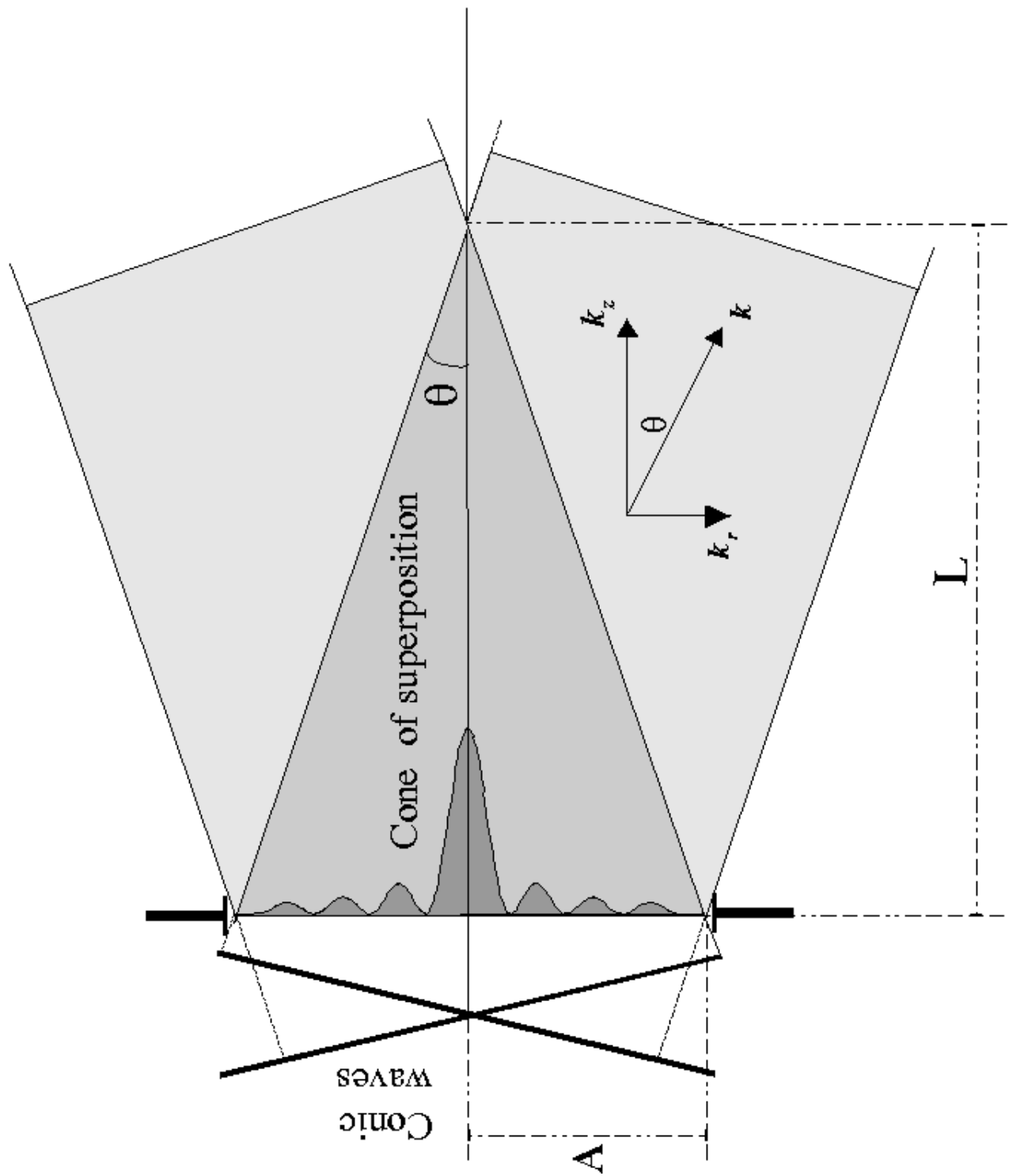
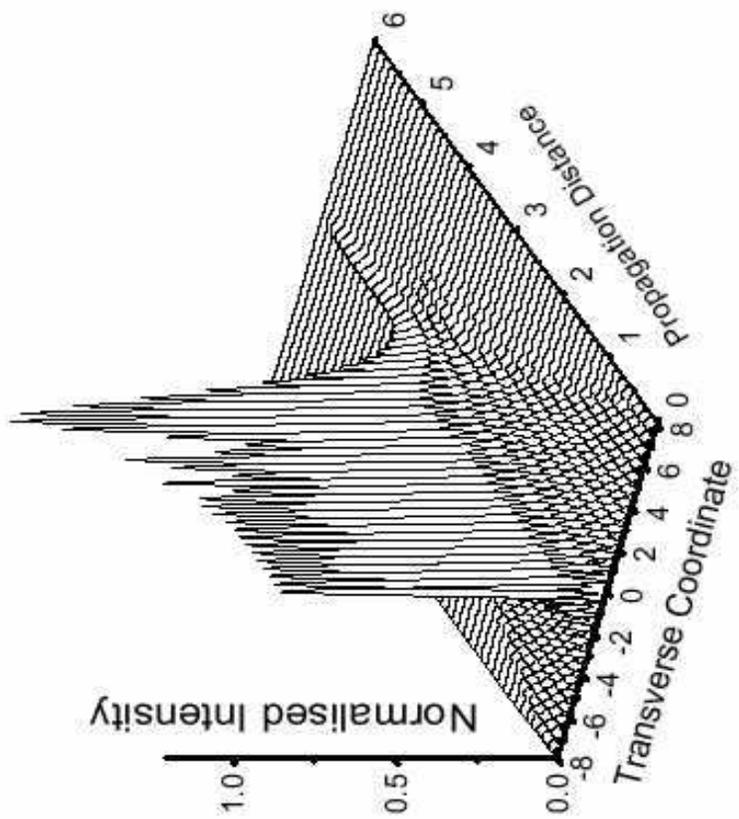
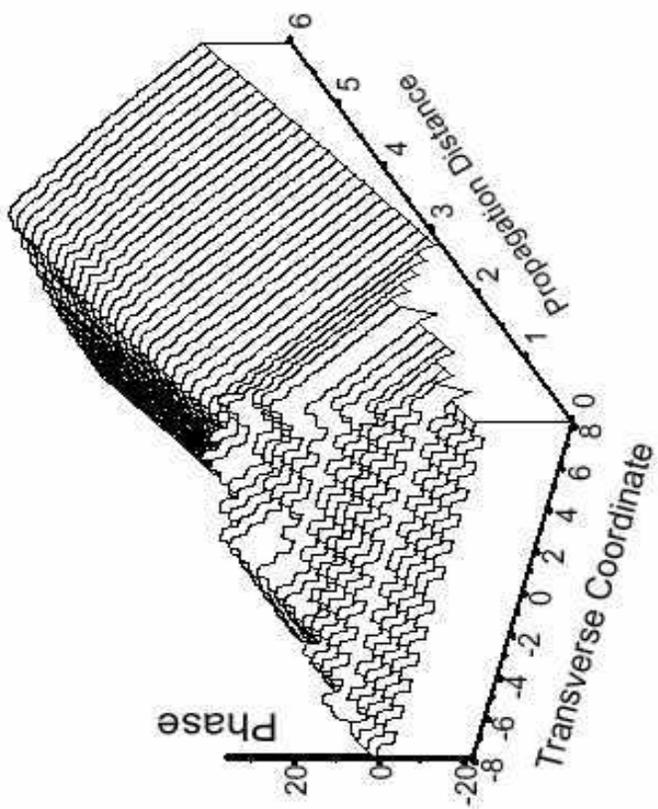
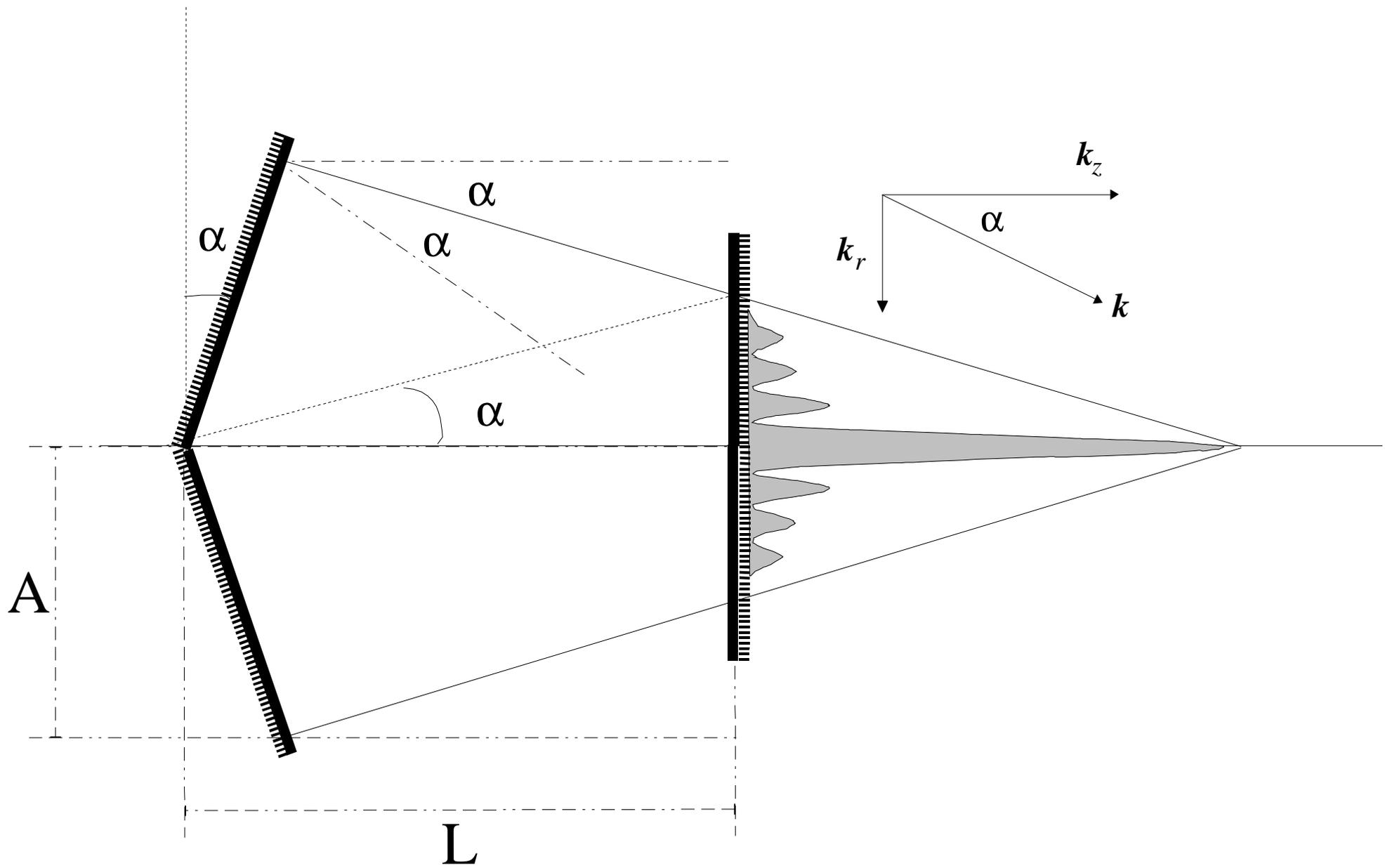
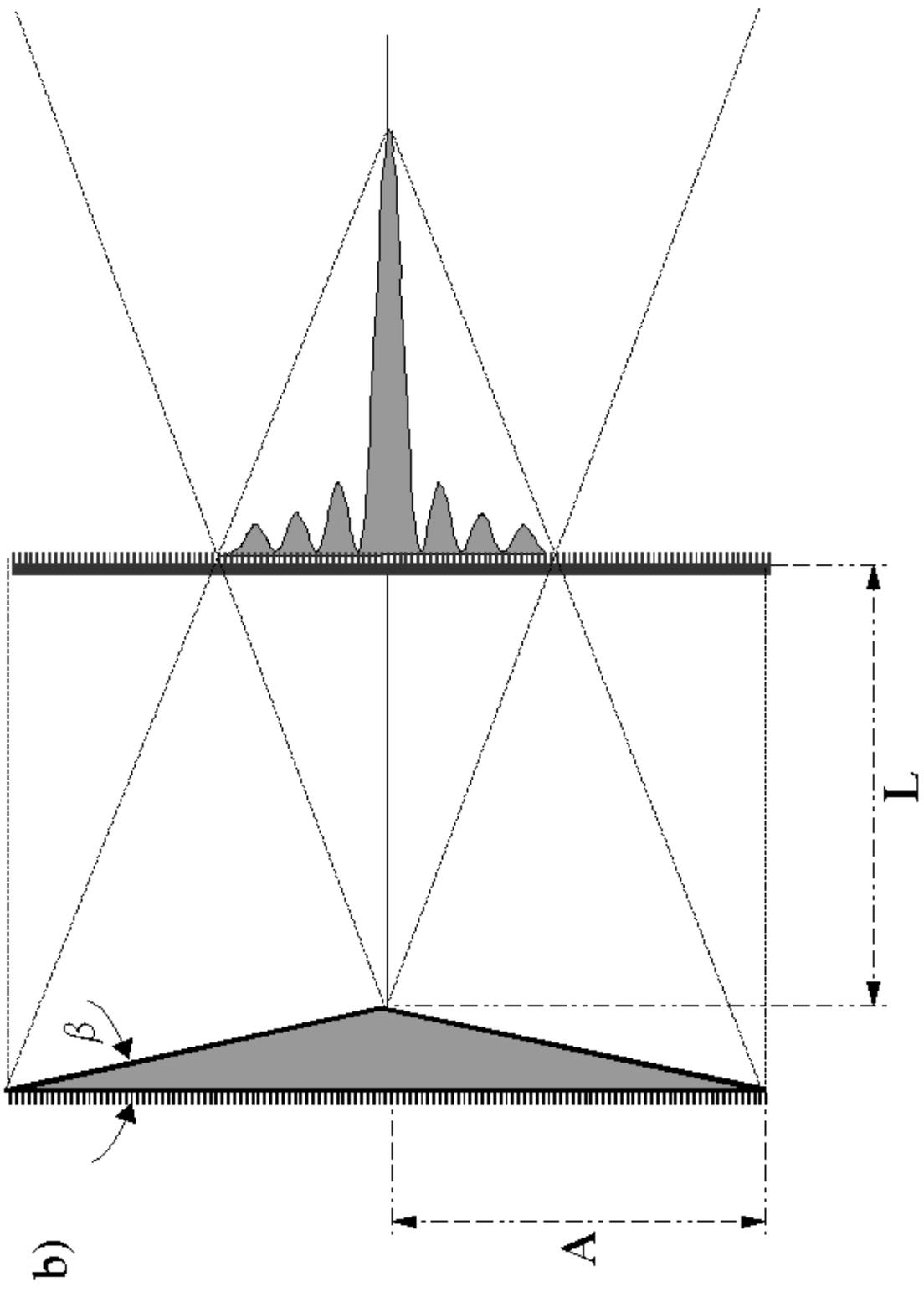
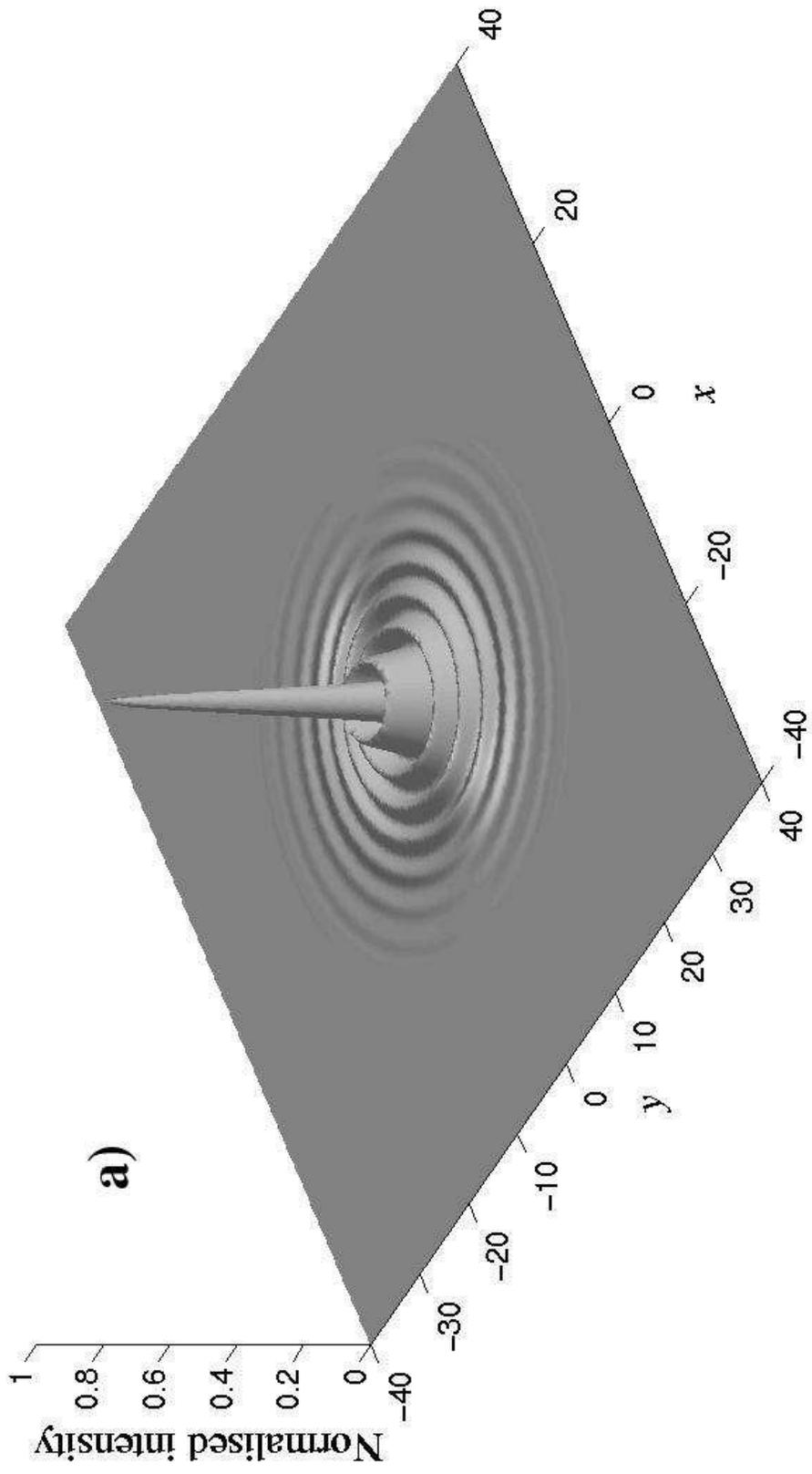
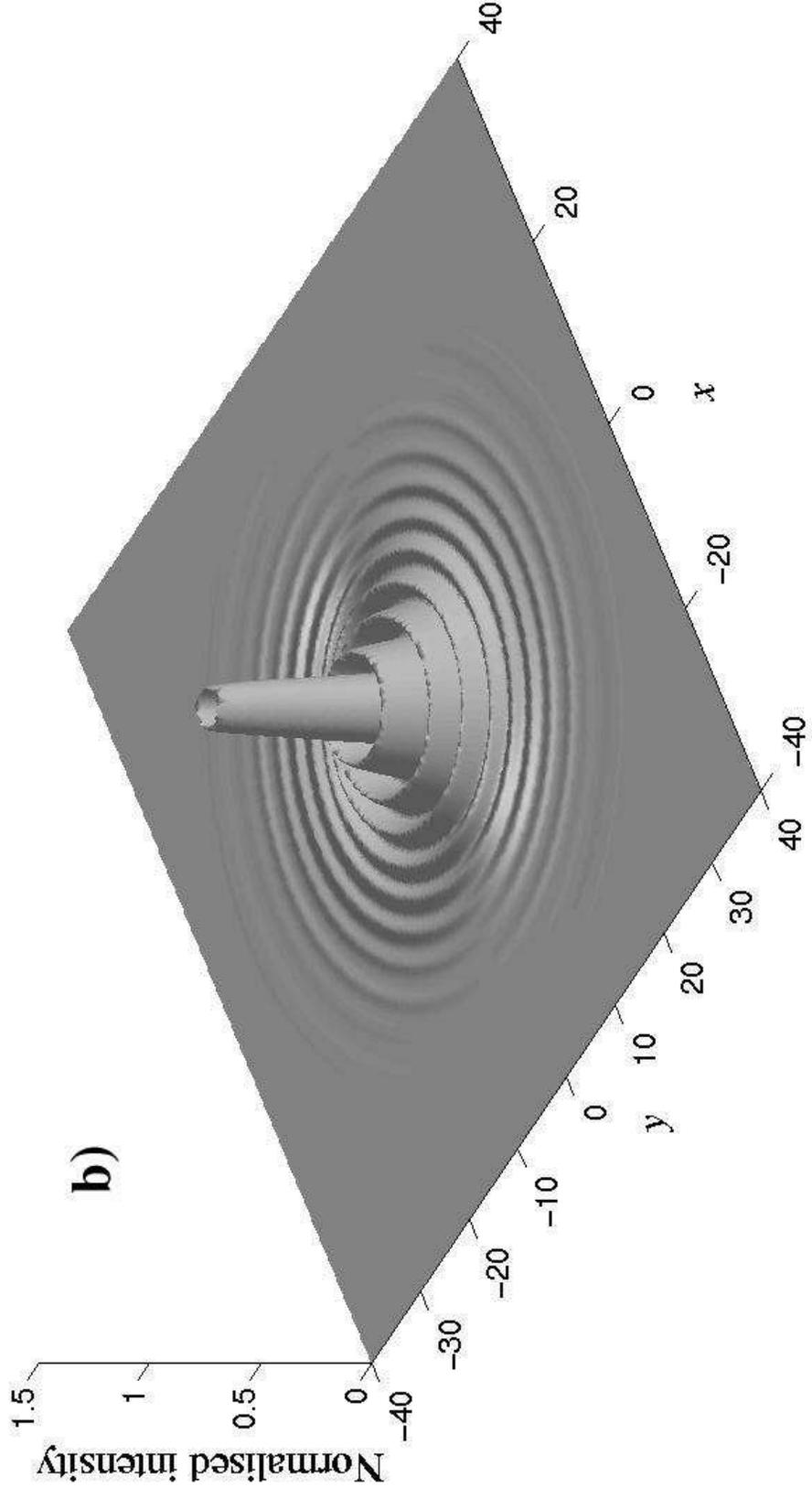
\end{document}